\documentclass[fleqn,usenatbib]{mnras}

\usepackage{newtxtext,newtxmath}

\usepackage[T1]{fontenc}

\DeclareRobustCommand{\VAN}[3]{#2}
\let\VANthebibliography\thebibliography
\def\thebibliography{\DeclareRobustCommand{\VAN}[3]{##3}\VANthebibliography}

\usepackage{graphicx}	
\usepackage{amsmath}	
\usepackage{color}
\usepackage{enumitem}
\usepackage{hyperref}
\usepackage{verbatim}
\usepackage{mathrsfs}
\usepackage[all]{hypcap} 
\usepackage{afterpage}    
\usepackage{marginnote}
\usepackage{multirow}
\usepackage{lineno}
\usepackage{float}
\usepackage{siunitx}

\title[Magnetic Field towards the Supernovae Remnant W44]{Velocity Gradients: Magnetic Field Tomography towards the Supernova Remnant W44}

\author[Liu, Hu \& Lazarian]{
Mingrui Liu$^{1}$,
Yue Hu$^{1,2}$\thanks{E-mail: yue.hu@wisc.edu},
A. Lazarian$^{2,3}$\thanks{E-mail: alazarian@facstaff.wisc.edu}
\\
$^{1}$Department of Physics, University of Wisconsin-Madison, Madison, WI, 53706, USA\\
$^{2}$Department of Astronomy, University of Wisconsin-Madison, Madison, WI, 53706, USA\\
$^3$Centro de Investigación en Astronomía, Universidad Bernardo O’Higgins, Santiago, General Gana 1760, 8370993,
Chile\\
}

\date{Accepted XXX. Received YYY; in original form ZZZ}

\pubyear{2020}

\begin{document}
\label{firstpage}
\pagerange{\pageref{firstpage}--\pageref{lastpage}}
\maketitle

\begin{abstract}
As a novel approach for tracing interstellar magnetic fields, the Velocity Gradient Technique (VGT) has been proven to be effective for probing magnetic fields in the diffuse interstellar medium (ISM). In this work, we verify the VGT in a broader context by applying the technique to a molecular cloud interacting with the supernovae remnant (SNR) W44. We probe the magnetic fields with the VGT using CO, $\rm HCO^+$, and H I emission lines and make comparison with the Planck 353 GHZ dust polarization. We show that the VGT gives an accurate measurement that coheres with the Planck polarization especially in intense molecular gas emission regions. We further study the foreground's contribution on the polarization that results in misalignment between the VGT and the Planck measurements in low-intensity molecular gas areas. We advance the VGT to achieve magnetic field tomography by decomposing the W44 into various velocity components. We show that W44's velocity component at $v\sim45$ km s$^{-1}$ exhibits the largest coverage and gives best agreement with Planck polarization in terms of magnetic field orientation.
\end{abstract}




\section{Introduction} 
\label{sec:intro}
Magnetism is of enormous significance in current astronomy studies. Magnetized turbulence in diffuse interstellar media (ISM;  \citealt{1981MNRAS.194..809L,1995ApJ...443..209A,2004ARAA..42..211E,2004RvMP...76..125M,2007ARA&A..45..565M}) are essential to critical astrophysics problems, including star formation (see \citealt{1956MNRAS.116..503M,1959ApJ...129..243S,2004RvMP...76..125M,2007ARA&A..45..565M,2020ApJ...897..123H}), cosmic ray propagation as well as acceleration \citep{1966ApJ...146..480J,1978MNRAS.182..147B,2002PhRvL..89B1102Y,2014ApJ...783...91C}. Turbulence-mediated reconnection is important for release of magnetic energy in astrophysical evnironments, the solar flares being one of the most conspicuous examples (see \citealt{2005AdSpR..35.1707K,2020PhPl...27a2305L}). The studies for Supernovae Remnants' (SNRs) evolution rely crucially on magnetism and turbulence as well. 

However, probing magnetic fields in SNRs confronts technical difficulties. Conventionally recognized techniques including starlight polarization in interstellar dust (see \citealt{2000AJ....119..923H}; \citealt{2012ApJS..200...20C,2012ApJS..200...21C,2013AAS...22135215C}) and the polarized thermal emission of aligned dust (see \citealt{2007JQSRT.106..225L}; \citealt{2015ARA&A..53..501A}) are limited by multiple factors in deep space observations. For one, the contribution from the polarized galactic foreground obscures the local plane-of-the-sky (POS) magnetic fields of the SNRs. Additionally, the POS magnetic fields can be revealed by the polarized synchrotron emission (\citealt{2016A&A...594A..25P}). It is adopted by \cite{2008A&A...482..783X, 2009A&A...503..827X} to trace the magnetic field on SNR G65.2 and S147 at wavelengths $\lambda\sim11$ cm and $\lambda\sim6$ cm. Nevertheless, the synchrotron polarization is contaminated by the effect of Faraday rotation, in particular at a long wavelength. Therefore, there is a demand for alternative methods for probing magnetic fields. 


The emergence of the Velocity Gradient Technique (VGT; \citealt{2017ApJ...835...41G};  \citealt{2017ApJ...837L..24Y}; \citealt{2018ApJ...853...96L}; \citealt{2018MNRAS.480.1333H}) provides a novel vision for interstellar magnetic field tracing. The technique roots in the properties of the magnetohydrodynamic (MHD) turbulence (\citealt{1995ApJ...438..763G}), and fast turbulent reconnection (\citealt{1999ApJ...517..700L}). Due to the reconnection, MHD turbulent eddies are stretched along local magnetic fields that percolate the eddies and thus become anisotropic (\citealt{1999ApJ...517..700L}). The resulting anisotropy induces gradients of velocity fluctuations' amplitude to be perpendicular to the directions of local magnetic fields, which enables the magnetic field tracing employing the amplitude of the turbulence's velocity fluctuations that are available from observations. 

Multiple observational works have verified the VGT to be reliable in magnetic field tracing. The technique proved to be reliable in probing magnetic fields multi-phase ISM, including diffuse neutral hydrogen regions \citep{2017ApJ...837L..24Y,2019ApJ...874...25G,2020ApJ...888...96H,2020MNRAS.496.2868L}, nearby molecular clouds (\citealt{2019NatAs...3..776H, 2019ApJ...884..137H}, \citealt{2020arXiv200715344A}), and the central molecular zone (\citealt{2021arXiv210503605H}). Previous practices have also suggested several potentials of the VGT, including eliminating the foreground interference \citep{2020ApJ...888...96H,2020MNRAS.496.2868L}. The VGT is also promising in decomposing the magnetic field of a molecular cloud into various components as a function of the line-of-sight (LOS) velocity. Consequently, it results in a three-dimensional (3D) magnetic field tracing.

In this research, we study the VGT’s potentials as discussed above by applying the technique to the SNR W44. The W44 is a Type II SNR locating 3 kpc away from the Sun, near $l = +33.^\circ 750$, $b = -1.^\circ 509$. The estimated age of the remnant is $(0.65–2.5) \times 10^4$ years (\citealt{2013ApJ...774...10S}; \citealt{1985MNRAS.217...99S}; \citealt{1997ApJ...488..781H}). A massive molecular cloud closely related to W44 has been observed in the region, which shows distinctive velocity components that provide ideal conditions for testing the VGT. We probe the magnetic fields of the SNR molecular cloud using CO (3-2) and $\rm HCO^+$ (1-0) emission lines (\citealt{2013ApJ...774...10S}) and make a comparison with the magnetic field map inferred from the Planck 353 GHz polarized dust emission (\citealt{2020A&A...641A..12P}). By combining with the neutral hydrogen data from the GALFA-H I survey (\citealt{2018ApJS..234....2P}), we test the VGT's capability in removing the foreground's contribution. Furthermore, we make the synergy of the VGT and the SCOUSEPY software to decompose the 3D magnetic fields of the CO (3-2) emission line.

The following contents are structured as such: in Sec.~\ref{sec:data}, we introduce the essential data from multiple observations that have contributed significantly to this research. In Sec.~\ref{sec:method}, we present details on the methodology for our work, including an explanation of the VGT and the SCOUSEPY software that is critical to the study. Sec.\ref{sec:results} is assigned to presentation of the results. We also discuss further details and conclude in the final section.

\section{Observational data} 
\label{sec:data}
\subsection{CO and HCO+ emission lines}
The emission lines of CO (3-2) and $\rm HCO^+$ (1-0) are observed with the Atacama Submillimeter Telescope Experiment (ASTE) 10m telescope and the Nobeyama Radio Observatory (NRO) 45m radio telescope, respectively (\citealt{2013ApJ...774...10S}). 

The $\rm HCO^+$ (1-0) observation sets its reference position at ($l$, $b$) = ($+33.^\circ 750$, $-1.^\circ 509$), with the FWHM beam resolution of $\sim18''$ and the beam efficiency of around $\eta_{\rm MB}\approx44\%$. The emission line was observed at both a wide frequency band of 512 MHz and a narrow frequency band of 32 MHz. For this research, we use the narrow-band mode, which gives the velocity coverage of 1700 km/s, resulting in a velocity resolution of 0.1 km/s, with a $17'' \times 17'' \times 1$ km/s  grid resolution. The final map has an RMS noise level of 0.075 K (1$\sigma$) in the unit of brightness temperature $T_{\rm MB}$.

The CO survey by the ASTE at 345.795 GHz outputted a beam size of $22''$, with the beam efficiency of 0.6, set to ($l$, $b$) = ($+33.^\circ 750$, $-1.^\circ 509$) for reference as well. A regular grid of the data was given as $8.''5 \times 8.''5 \times 1.0$ km/s, corresponding to the velocity coverage of 444 km/s. The overall RMS noise level is 0.15 K (1$\sigma$) in $T_{\rm MB}$ . More details can be found in \cite{2013ApJ...774...10S}.

\subsection{GALFA-H I}
We also use the 21 cm emission line of neutral hydrogen from the Galactic Arecibo L-Band Feed Array H I (GALFA-H I) survey (\citealt{2011ApJS..194...20P}). The GALFA-H I survey has a velocity resolution of 0.18 km/s, a spatial beam resolution of around $4’$. A typical 1 km/s velocity channel in the data has an average RMS noise level of 85mK  (\citealt{2011ApJS..194...20P}). The data for our study are extracted from the second release of the GALFA survey in 2018 (\citealt{2018ApJS..234....2P}).

\subsection{Planck polarization}
To obtain the magnetic field orientation from polarized dust emission, we adopt the data from the 2018 publication of the Planck project (Planck 3rd Public Data Release 2018 of High-Frequency Instrument; \citealt{2020A&A...641A..12P}) for this work. The magnetic field orientation is defined as $\phi_{\rm B}=\phi+\pi/2$, inferred from the polarization angle $\phi$ along with the polarization fraction $p$, specifically:
        \begin{equation}
        \begin{aligned}
           \phi&=\frac{1}{2}\arctan(-U,Q)\\
           p&=\sqrt{Q^2+U^2}/I
        \end{aligned}
        \end{equation}
where $I$, $Q$, and $U$ refer to the intensity of dust emission and Stokes parameters, respectively. $-U$ resolves the angle conversion from the HEALPix system to the IAU system, and the multi-variables arc-tangent function denotes the radian periodicity. We minimize the noise on the maps by smoothing the results with a Gaussian filter, regularizing the nominal angular resolution of the data to $10'$. 

\section{Methodology}
\label{sec:method}
\subsection{The Velocity Gradient Technique}
\subsubsection{Theoretical consideration}
The VGT sets its basis from the advanced MHD turbulence (\citealt{1995ApJ...438..763G},  noted as GS95 hereafter) as well as the fast turbulent reconnection theory (\citealt{1999ApJ...517..700L}, noted as LV99 hereafter). As introduced before, the MHD turbulent eddies exhibit anisotropy as they are elongating along the magnetic field, proposed by the GS95 study. The anisotropy was derived with regard to the "critical balance" condition, where the cascading time ($k_{\perp}v_l)^{-1}$ is equal to the wave period ($k_{\parallel}v_{\rm A})^{-1}$, as $k_{\perp}$ and $k_{\parallel}$ denote wavevectors perpendicular and parallel to the magnetic field direction respectively. Turbulent velocity $v_l \propto l^{1/3}$ at scale $l$ accounts for the Kolmogorov-type turbulence and $v_{\rm A}$ is the Alfv\'en velocity. Consequently, the quantitative expression of the MHD turbulent anisotropy in Fourier space is derived as: 
\begin{equation}
\begin{aligned}
          \textit{$k_{\parallel}$}  \propto   \textit{$k_{\perp}^{2/3}$}
\end{aligned}
\end{equation}
The GS95 anisotropy scaling is derived in the global reference frame, which is defined by the mean magnetic field. However, only the anisotropy that is independent of the motion scale will be observed in this frame, as small eddies are blanked out by larger eddies with the most significant contributions to the whole picture (\citealt{2000ApJ...539..273C}).

Consider the fact that, as discussed in LV99, magnetic fields give minimal resistance to the motions of eddies perpendicular to the local direction of the magnetic field. Thus, the eddies obey the hydrodynamic Kolmogorov law $v_{l,\bot} \propto l^{1/3}_\bot$, where $v_{l,\bot}$ is the turbulence’s velocity perpendicular to the local magnetic field at scale $l$. The anisotropy scaling in the local reference frame, which is defined by the local magnetic field, was later derived in LV99:
\begin{equation}
\begin{aligned}
l_{\parallel}  =  L_{\rm inj}(\frac{l_{\bot}}{L_{\rm inj}})^{2/3}M_{\rm A}^{-4/3}, M_{\rm A} \le 1
\end{aligned}
\end{equation}
where $l_{\parallel}$ denotes the scale parallel to the local magnetic field, $L_{\rm inj}$ denotes the turbulent injection scale; the Alfvén-Mach number is $M_{\rm A}$, and $l_{\bot}$ indicates the scale perpendicular to the local magnetic field. Furthermore, the scaling for the anisotropy towards the velocity fluctuation in the turbulent eddies as well as its gradient is given as (LV99; \citealt{2020ApJ...897..123H}):
\begin{equation}
\label{eq.4}
        \begin{aligned}
           &v_{l,\bot}  =  v_{\rm inj}(\frac{l_{\bot}}{L_{\rm inj}})^{1/3}M_{\rm A}^{1/3}\\
           &\nabla v_l\propto \frac{v_{l,\bot}}{l_{\bot}}\approx\frac{v_{\rm inj}}{L_{\rm inj}}M_{\rm A}^{1/3}(\frac{l_{\bot}}{L_{\rm inj}})^{-2/3}
        \end{aligned}
        \end{equation}
where $v_{\rm inj}$ corresponds to the injection velocity. As an indicator of anisotropy, the gradient points to the direction with the maximum variation of velocity fluctuation's amplitude, which is perpendicular to the local magnetic field direction. 

The discussion of MHD turbulence above deals with sub-Alv\'enic turbulence, i.e., the injection velocity $v_{\rm inj}$ of turbulence is less than the Alfv\'en velocity $v_{\rm A}$. For SNRs, the hydro effect is significant and the turbulence is usually super-Alfv\'enic, i.e., with $M_{\rm A}=v_{\rm inj}/v_{\rm A}>1$. For such turbulence, the motions at the injection scale are hydrodynamic due to the relatively weak back-reaction of the magnetic field. However, as the kinetic energy of turbulent motions follows the nearly isotropic Kolmogorov cascade, i.e., $v_l\sim l^{1/3}$, the importance of magnetic back-reaction gets stronger at smaller scales. Eventually, at the scale $l_{\rm A} = L_{\rm inj}M_{\rm A}^{-3}$, the turbulent velocity becomes equal to the Alfv\'en velocity (\citealt{2006ApJ...645L..25L}) showing the GS95 anisotropic scaling relation.

\subsubsection{The Recipe of VGT}
For this work, we employ specifically the Velocity Channel Gradients Technique (VChGs) as the primary analytical approach. The essence of the VChGs is the thin velocity channel, mapped as  Ch(x,y). The statistics of the intensity fluctuations in PPV space and their relations to the underlying statistics of turbulent velocity and density were firstly proposed by \cite{2000ApJ...537..720L}. It suggests that velocity fluctuations dominate over density fluctuation when the channel width is sufficiently thin. Specifically, a thin channel is defined as the following criteria:
\begin{equation}
    \begin{aligned}
        &\Delta v < \sqrt{\delta (v^2)}, \text{thin channel}\\
         &\Delta v \geq \sqrt{\delta (v^2)}, \text{thick channel}
        \end{aligned}
        \end{equation}
where $\Delta v$ is the channel width and $\sqrt{\delta (v^2)}$ represents velocity dispersion calculated from velocity centroid. Accordingly, the VChGs is calculated from:
\begin{equation}
        \begin{aligned}
           \nabla{\rm_ x} {\rm Ch_i} (x,y) =  \textit{$G_{\rm x}$} * \textit{${\rm Ch_i} (x,y)$} \\
           \nabla_{\rm y} {\rm Ch_i} (x,y)  =  \textit{$G_{\rm y}$} * \textit{${\rm Ch_i} (x,y)$}
        \end{aligned}
        \end{equation}
where $\bigtriangledown_{\rm x} {\rm Ch_i} (x,y)$, $\bigtriangledown_{\rm y} {\rm Ch_i} (x,y)$ denote the x, y components of the gradients in the thin channel map; the asterisks indicate the convolutions with Sobel kernels $G_{\rm x}$, $G_{\rm y}$ for probing pixelized gradient map $\psi_{\rm g}^{\rm i} (x,y)$, namely
\begin{equation}
       \begin{aligned}
&\psi{\rm ^i_g} (x,y)=\tan^{-1}(\frac{\nabla{}{}_{\rm y} {\rm Ch_i} (x,y)}{\nabla_{\rm x} {\rm Ch_i} (x,y)})
\end{aligned}
\end{equation}

In this paper we follow the procedure of sub-block averaging introduced in \cite{2017ApJ...837L..24Y}. The map is divided into sub-blocks and the Gaussian fitting to individual sub-blocks to derive the most probable orientation of the gradients of the sub-blocks. For each thin velocity channel, such an averaging method is employed; consequently, we acquire compiled gradient maps $\psi{\rm _{g s}^i} (x,y)$ of the same number as the thin velocity channel ($n_{\rm v}$). The pseudo Stokes parameters, similar to the essences of the Planck polarization-probed magnetic fields, are introduced:
\begin{equation}
        \begin{aligned}
           &Q_{\rm g} (x,y)  =  \sum^{n_{\rm v}}_{{\rm i=1}} {\rm Ch_i} (x,y) \cos(2\psi{\rm ^i_{gs}} (x,y))\\
           &U_{\rm g} (x,y)  =  \sum^{n_{\rm v}}_{{\rm i=1}} {\rm Ch_i} (x,y) \sin(2\psi{\rm ^i_{gs}} (x,y))\\
           &\psi_{\rm g} = \frac{1}{2}\tan^{-1}(\frac{U_{\rm g}}{Q_{\rm g}})
        \end{aligned}
        \end{equation}
where $\psi_{\rm g}$ denotes the pseudo polarization angle, and thus the POS magnetic field direction is defined as $\psi_{\rm B} = \psi_{\rm g} + \pi/2$, as the same statistics introduced in Sec. 2 about the Planck polarization. 

The relative orientation between the two magnetic field directions probed with the Planck polarization ($\phi_B$) and the gradients ($\psi_B$) is measured with the Alignment Measure (AM; \citealt{2017ApJ...835...41G}), defined as
\begin{equation}
        \begin{aligned}
           {\rm AM}  =  2(\cos^2(\theta_{\rm r})-\frac{1}{2})
        \end{aligned}
\end{equation}
the theta angle is defined as $\theta_{\rm r} = \vert\phi_{\rm B}-\psi_{\rm B}\vert$. AM is a relative scale ranging from -1 to 1, with AM = 1 indicating that $\phi_{\rm B}$ and $\psi_{\rm B}$ are parallel, and AM = -1 denoting that the two are orthogonal.  

\begin{figure*}
\label{fig.intensity}
    \centering
    \includegraphics[width=1.0\linewidth]{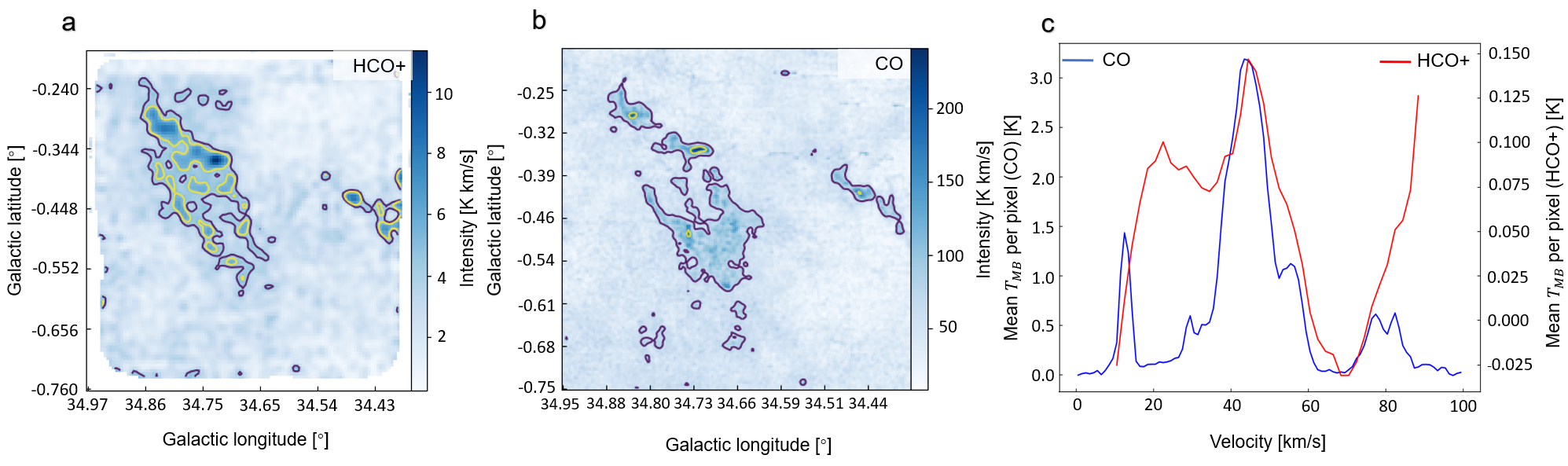}
    \caption{\textbf{Panel a \& b}: the integrated intensity maps of the $\rm HCO^+$ (1-0) emission line (left) and the $\rm CO$ (3-2) line (right). The purple contour lines in the intensity maps start from the value of mean intensity. \textbf{Panel c:}
    the intensity-velocity spectrum of $\rm HCO^+$ (1-0; red) and $\rm CO$ (3-2; blue). }
\end{figure*}

\begin{table*}
\centering
\label{tab:param}
\begin{tabular}{| c | c | c | c |}
\hline
Physical Parameter & Symbol/Definition & Value & Reference\\ \hline \hline
Velocity dispersion of shocked gas & $v_{\rm s}$ & 6.8 km $\rm s^{-1}$ &  \cite{2013ApJ...774...10S}\\  
Velocity dispersion of quiescent gas & $v_{\rm q}$ & 2.3 km $\rm s^{-1}$ &  \cite{2013ApJ...774...10S}\\
$\rm H_2$ Volume number density & $n_0$ & 100 - 300 $\rm cm^{-3}$ &  \cite{2011ApJ...742L..30G} \& \cite{2014AA...565A..74C}\\
Magnetic field strength & $B$ & 70 ${\rm \mu G}$ & \cite{2011ApJ...742L..30G}\\
Mass of an H atom & $m_{\rm H}$ & 1.67$\times10^{-24}$ g & \cite{2008AA...487..993K} \\
Mean molecular weight & $\mu_{\rm H_2}$ & 2.8 & \cite{2008AA...487..993K}\\
Injection scale  & $L_{\rm inj}$ & 100.00 pc & \cite{2004ARAA..42..211E}\\
3D turbulent velocity& $v_{\rm tur}=\sqrt{3(v_{\rm s}^2-v_{\rm q}^2)}$ & 11.08 km $\rm s^{-1}$ & Derived\\
Volume mass density & $\rho_0=n_0\mu_{\rm H_2} m_{\rm H}$ & 3.34 $\times10^{-23}$ g $\rm cm^{-3}$ & Derived \\
Alfv\'en velocity & $v_{\rm A}=B/\sqrt{4\pi\rho_0}$ & 5.27 - 9.13 km $\rm s^{-1}$ & Derived\\
Alfv\'en Mach number & $M_{\rm A}=v_{\rm tur}/v_{\rm A}$ & 1.21 - 2.10& Derived\\
Transition scale  & $l_{\rm A}=L_{\rm inj}M_{\rm A}^{-3}$ & 10.80 - 56.00 pc & Derived\\
Beam resolution of CO & $l_{\rm CO}$ & 0.26 pc & Derived\\
Beam resolution of HCO$^+$ & $l_{\rm HCO^+}$ & 0.32 pc & Derived\\
\hline
\end{tabular}
\caption{Physical parameters of the W44 SNRs. 
}
\end{table*}

\subsection{SCOUSEPY}
To explore the VGT's potential in probing three dimensional POS magnetic fields,
we use the SCOUSEPY, a Python software based on the Semi-automated multi-COmponent Universal Spectral-line fitting Engine (SCOUSE) proposed by \cite{2016MNRAS.457.2675H,2019MNRAS.485.2457H}, to decompose the spectrum to individual Gaussian components. 

SCOUSEPY is designed for fitting a large number of spectroscopic data with a multi-stages procedure. The software firstly divides the spatial data of the user-selected region into small areas, named Spectral Averaging Areas (SAAs), and outputs a spectrum averaged spatially for each of the SAAs. Users manually mark the intended spectral range of the data, then SCOUSEPY performs the fitting procedures, which adopts the appropriate fitting algorithms from multiple fitting profiles, including Gaussian, Voigt, Lorentzian, and hyperfine structure fitting. For our work in the SNR, SCOUSEPY chooses the Gaussian method. The final result consists of the best-fitting solutions of pixels and velocity components for the data. More details can be found in \cite{2016MNRAS.457.2675H}.

In this work, we obtain four decomposed Gaussian components of W44's CO emission lines with SCOUSEPY. We follow the tolerance criteria proposed in \cite{2016MNRAS.457.2675H}: (i) all detected
components must have a brightness temperature greater than three
times the local noise level; (ii) each Gaussian component must have an FWHM line-width of at least one channel: (iii) for two Gaussian components to be considered distinguishable, they must be separated by at least half of the FWHM of the narrowest of the two. (iv) the size of SAAs is set to six.

\begin{figure*}
    \centering
    \includegraphics[width=1.0\linewidth]{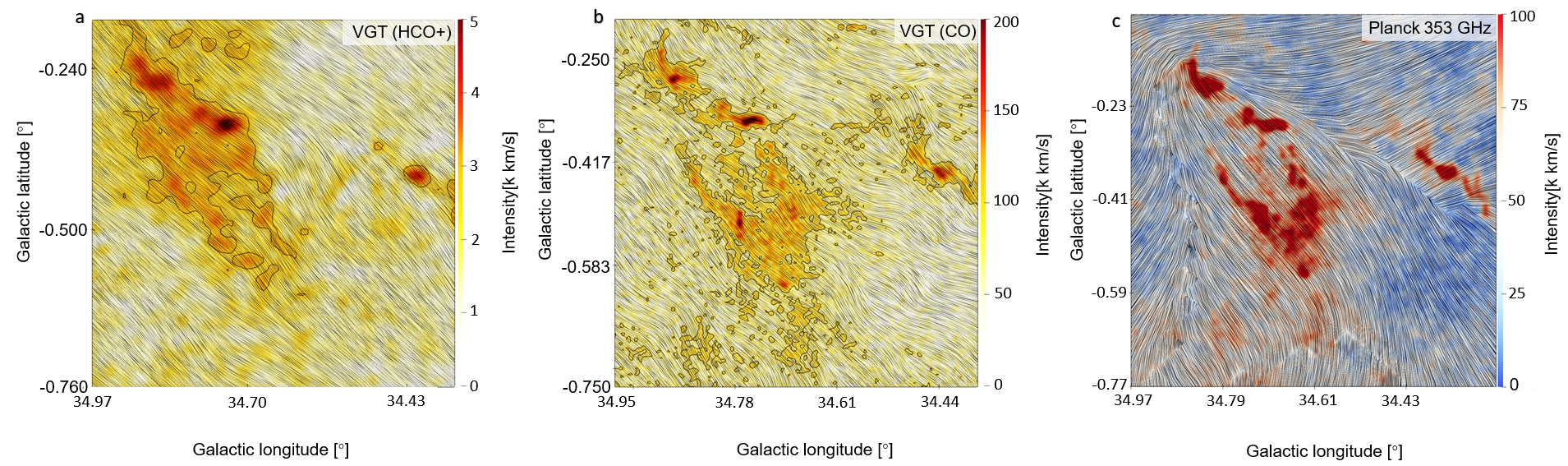}
    \caption{Visualization of magnetic fields towards the W44 SNRs using the Line Integral Convolution (LIC). The magnetic fields were inferred from Planck 353 GHz polarized dust emission (panel c) and VGT using $\rm CO$ (panel b) and $\rm HCO^+$ (panel a) emissions. VGT measurements are overlaid on corresponding integrated intensity color maps.}
    \label{fig.vgt}
\end{figure*}

\section{Results}
\label{sec:results}
\subsection{The POS magnetic fields in W44}

\begin{figure*}
\label{fig.HI-Planck}
    \centering
    \includegraphics[width=1\linewidth]{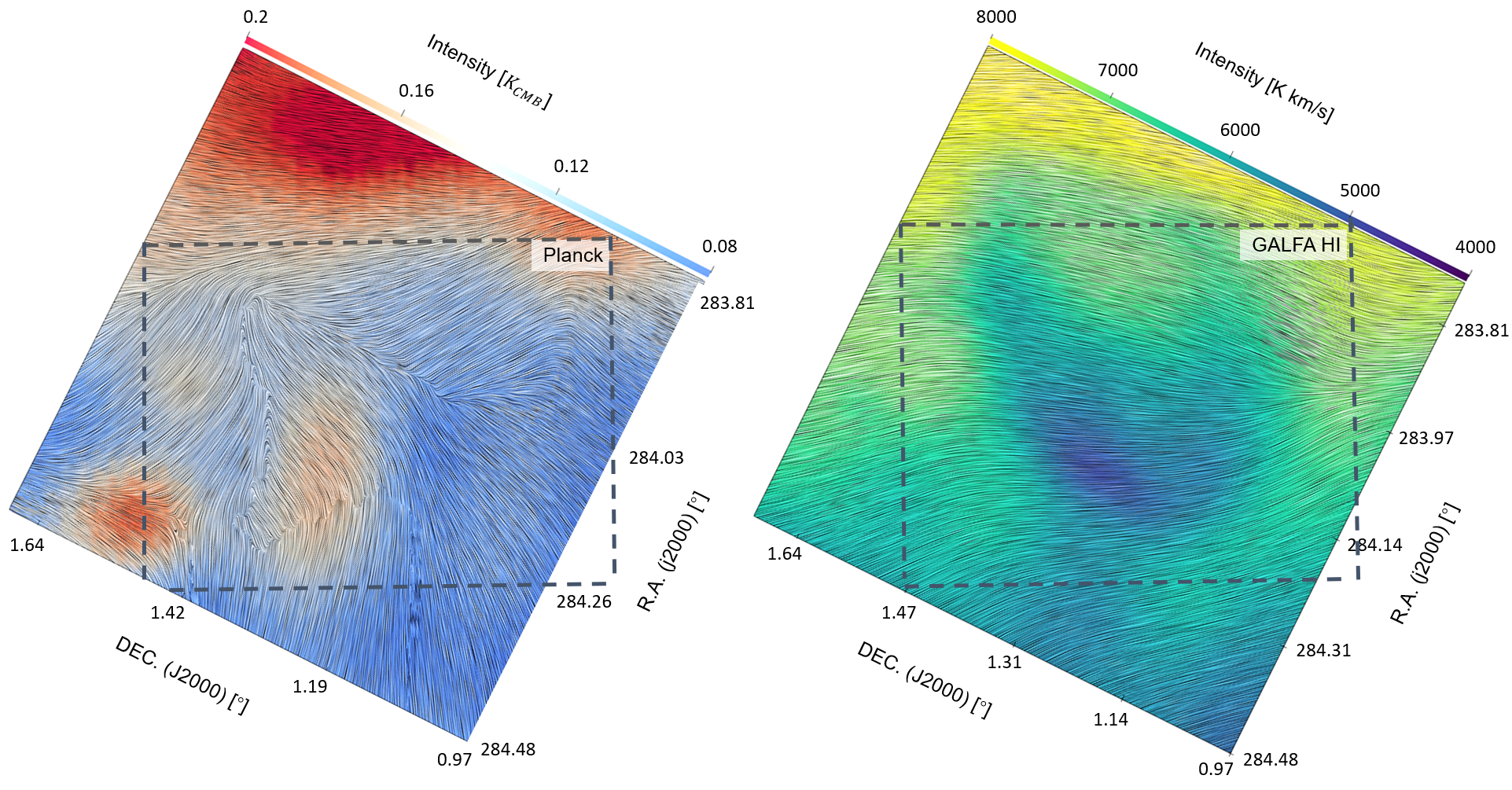}
    \caption{Visualizations of the magnetic field orientation of the SNR foreground region probed with the Planck polarization ($\phi_{\rm B}^{\rm H I}$, left panel) as well as with the VGT on the GALFA H I emission ($\psi_{\rm B}^{\rm H I}$, right panel). The graphs have been rotated ($\sim63^\circ$) to convert the original equatorial coordinates to the galactic coordinates.  The grey dashed-lined boxes mark the approximate area corresponding to the same region used in Fig.~\ref{fig.vgt} in the galactic coordinates. 
    }
\end{figure*}

\begin{figure*}
\label{fig.colden_HI}
    \centering
    \includegraphics[width=1\linewidth, height=0.36\linewidth]{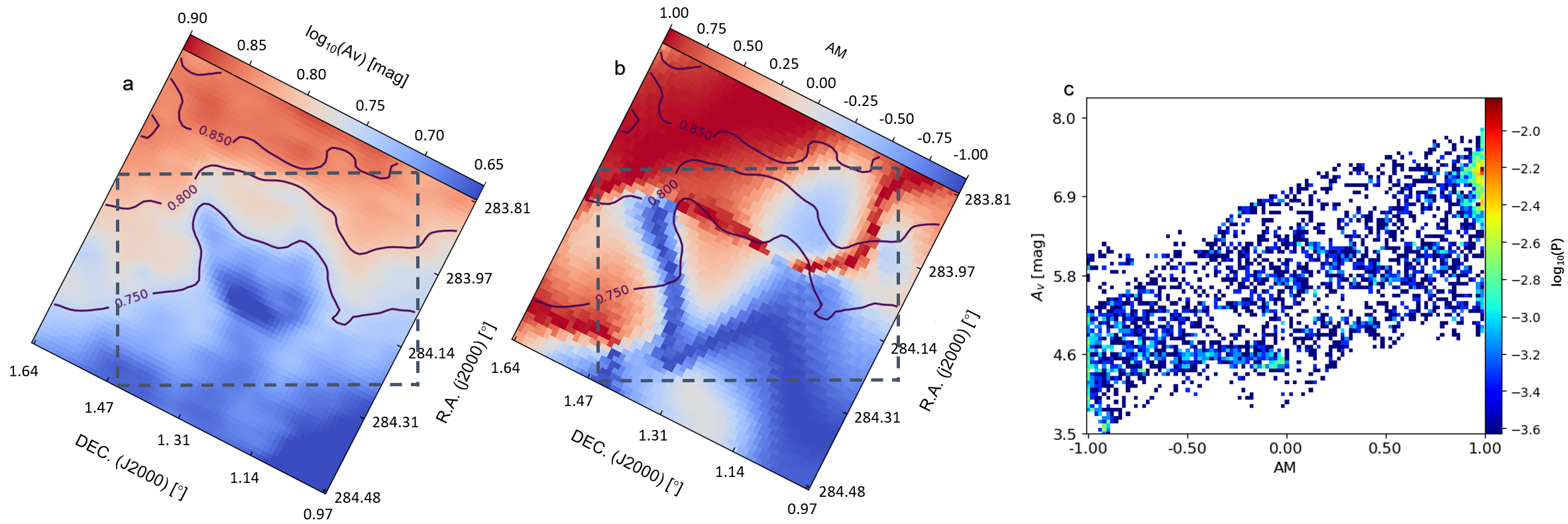}
    \caption{\textbf{Panel a}: the extinction map of the foreground H I emission. \textbf{Panel b}: the distribution of AM between the magnetic field orientations obtained by the H I emission ($\phi_{\rm B}^{\rm H I}$) and the Planck polarization ($\psi_{\rm B}^{\rm H I}$). The purple contour lines in both maps mark the order of magnitudes of the optical extinction $A_{\rm v}$. The graphs have been rotated ($\sim63^\circ$) to convert the original equatorial coordinates to the galactic coordinates. The dashed-lined boxes mark the area of used in Fig.~\ref{fig.vgt} in the galactic coordinates. \textbf{Panel c}: the $A_{\rm v}$-AM histogram of the H I emission. P in the colorbar label indicates the percent of sampled data points of the two variables.}
\end{figure*}

\begin{figure*}
\label{fig.colden_CO}
    \centering
    \includegraphics[width=1\linewidth]{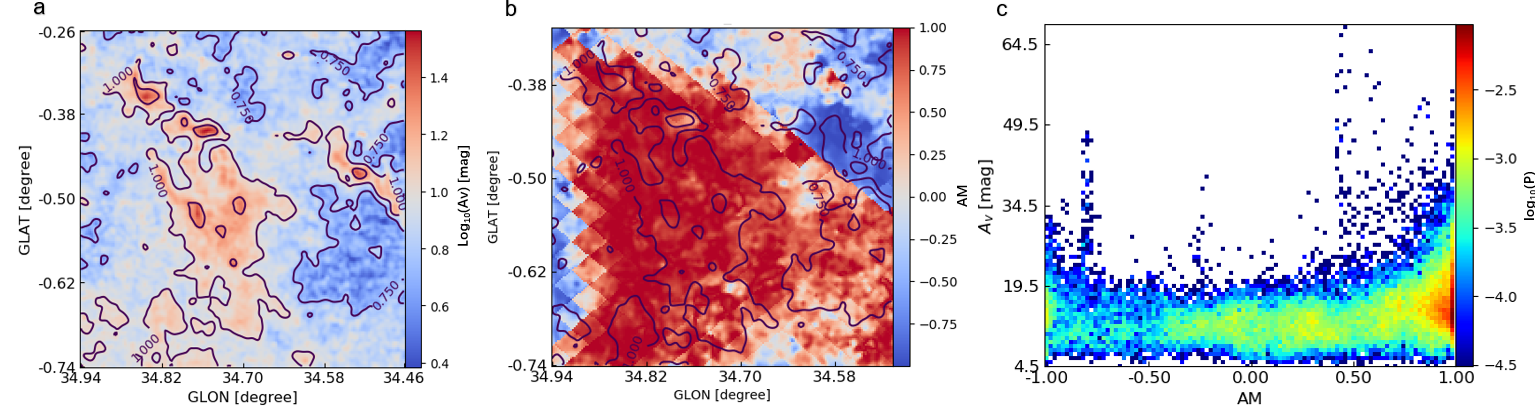}
    \caption{\textbf{Panel a}: the extinction map of the $\rm CO$ (3-2) emission. \textbf{Panel b}: the distribution of AM between the magnetic field orientations obtained by the VGT ($\phi_{\rm B}^{\rm CO}$) and the Planck polarization ($\psi_{\rm B}^{\rm CO}$). The purple contour lines in both maps mark the order of magnitudes of the optical extinction $A_{\rm v}$. \textbf{Panel c}: the $A_{\rm v}$-AM histogram of the $\rm CO$ (3-2) emission. P in the colorbar label indicates the percent of sampled data points of the two variables. }
\end{figure*}

Fig.~\ref{fig.intensity} displays the integrated intensity maps of the $\rm HCO^+$ (1-0) emission line and the $\rm CO$ (3-2) line, as well as the respective intensity-velocity spectrum. The contours outline the overall structure of the SNR molecular cloud above mean intensity. The molecular cloud appears as filamentary elongating along the east-north direction. In particular, the spectrum of the two gas lines indicate three apparent velocity components at $v\approx10$ km/s, $45$ km/s, and $80$ km/s. Note that the spectrum is averaged over the entire field-of-view shown in the panel a \& b. The narrow bandwidth of the original observation truncates $\rm HCO^+$ at velocity $\sim90$ km/s so that its information is incomplete. In the following, we focus the study on analyzing $\rm CO$.

As discussed in Sec.3, once the telescope resolves the SNRs at a scale smaller than $l_{\rm A}$, we can expect the turbulence to be anisotropic and velocity gradients are perpendicular to their local magnetic fields. To calculate $l_{\rm A}$, we use several physical parameters of W44 from literature, which is summarized in Tab~\ref{tab:param}. The emission lines towards W44 have beam resolution smaller than $l_{\rm A}\approx10.80 $ - $56.00$ pc. Therefore, the MHD turbulence in the SNR molecular cloud region is anisotropic, which fits the essential condition of applying the VGT.

We then obtain the POS magnetic field map of the same region by applying the VGT to both the CO and $\rm HCO^+$ emission data covering all velocity components. Pixels where the brightness temperature is less than three times the RMS noise level are blanked out. We average the gradients over each 20$\times$20 pixels sub-block and smooth the gradients map $\psi_g$ with a Gaussian filter $\rm FWHM\approx3'$. 

The results are displayed in Fig.~\ref{fig.vgt}. We can see that the magnetic fields inferred from both CO and $\rm HCO^+$ are elongating along the filamentary intensity structures, showing insignificant variation. Additionally, we make a comparison with the magnetic fields obtained from the Planck 353 GHz polarization data (see Fig.~\ref{fig.vgt}). Visually we can see that in high-intensity regions, the magnetic field orientations in both the VGT maps and the Planck maps agree with each other well. Notably, both measurements reveal the fact that W44 is elongating along the magnetic field direction. As the shocked gas is propagating in the direction perpendicular to the filamentary structures \citep{2013ApJ...774...10S}, our analysis suggests a picture of perpendicular shock, in which both the upstream and downstream plasma flows are perpendicular to the magnetic field, as well as the shock front. This indicates that the relative upstream plasma velocity, which is most likely contributed by SNR explosion, is greater than the downstream velocity so that the magnetic field is compressed. The remnant's expansion might introduce an extra velocity gradient, which is not associated with our theoretical picture of MHD turbulence. This type of velocity gradient, however, is expected to be dominated by the jump of velocity, as well as the acceleration of upstream/downstream plasma. Consequently, the velocity gradient of the perpendicular shock is still perpendicular to the magnetic fields. The agreement of VGT and Planck observed in Fig.~\ref{fig.vgt} are contributed by both velocity gradients of MHD turbulence and perpendicular shocks. Note that the amplitude of turbulence's gradient increases at small scales (see Eq.~\ref{eq.4}), but the shock's gradient does not. The observed velocity gradient here is, therefore, expected to be dominated by MHD turbulence.

In addition, in the low-intensity regions, in particular at the west-north corner, the magnetic fields inferred from the VGT significantly differ from the one acquired from the Planck polarization. Such a difference is further investigated in Fig.~\ref{fig.HI-Planck}, where we look into the contribution of the polarized dust foreground, that potentially results in the inconsistency between the VGT and the Planck in low-intensity regions.

\begin{figure*}
\label{fig.decomposed-spectrum}
    \centering
    \includegraphics[width=1\linewidth]{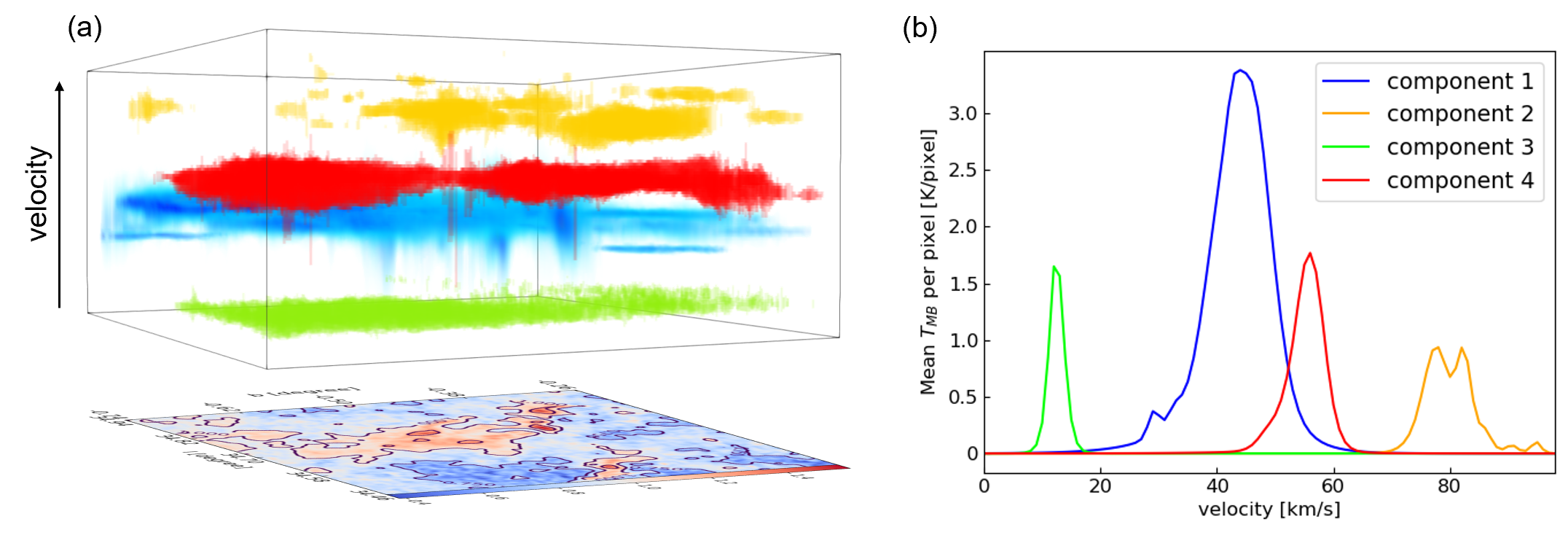}
    \caption{\textbf{Panel a}: visualization of the decomposed components of the W44 SNRs at various velocities. \textbf{Panel b}: the velocity-intensity spectrum of different  Gaussian components.}
\end{figure*}
\begin{figure*}
\label{fig.velocity-components}
    \centering
    \includegraphics[width=1.0\linewidth]{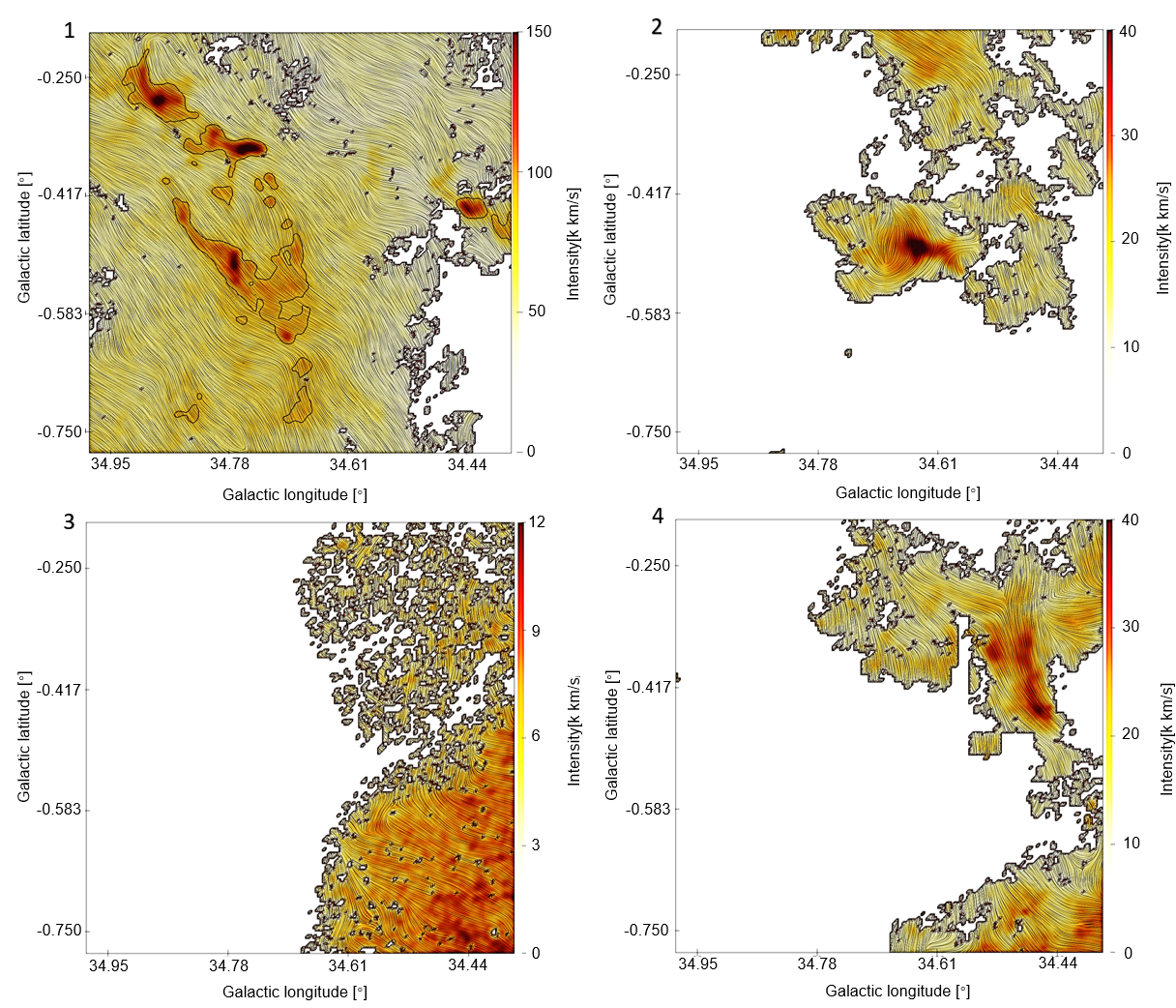}
    \caption{Visualizations of the four velocity-wise magnetic field components of the SNR W44, numbered as labeled. Magnetic fields are overlaid on corresponding integrated intensity color maps of $\rm CO$ emission.}
\end{figure*}

\subsection{Contribution from the foreground}


In Fig.~\ref{fig.vgt}, we observed misalignment between the magnetic field inferred from the VGT and Planck polarization. One possible contribution of the misalignment comes from the polarized dust foreground, as well as background. To find the contribution, we adopt the neutral hydrogen H I emission line from the GALFA-H I survey to trace the magnetic field through VGT. As H I is well mixed with dust in the foreground/background, VGT consequently results in a foreground/background magnetic field map. 

Fig.~\ref{fig.HI-Planck} presents the magnetic fields probed by VGT using H I, as well as probed by the Planck polarization. As previously performed, pixels with the brightness temperatures less than three times the RMS noise level are left blank; the gradients values are averaged over the 20$\times$20 pixels sub-blocks, and the gradients map $\psi_{\rm g}$ is smoothed with a $\rm FWHM\approx10'$ Gaussian filter. In the high-intensity regions, especially at the top-north corner, the magnetic field orientations in both maps cohere with each other well. The Planck data is dominated by the foreground interference as the H I signals become the most intense in the areas. Near the W44, the intensity of the H I emission diminishes to the minimum, so that the foreground's contribution to the Planck polarization is less significant.

We further quantitatively explore the foreground's contribution by calculating hydrogen column density as well as the optical extinction $A_{\rm v}$, which is proportional to the number of grains along the LOS, contributed by the W44 and the foreground/background. The hydrogen column density of the foreground/background is (\citealt{2019ApJ...872...56P}):
\begin{equation}
\label{eq.10}
    \begin{aligned}
   N_{\rm H^{f}}&=N_{\rm H I}+2N_{\rm H_2}\approx N_{\rm H I}=\int_{v_{\rm min}}^{v_{\rm max}} 1.823\times10^{18}T_{\rm MB}^{\rm HI} dv
    \end{aligned}
\end{equation}
where $N_{\rm HI}$ denotes the column density of \ion{H}{1} in the unit of $\rm{cm^{-2}}$, $N_{\rm H_2}$ symbols the column density of $\rm H_2$, and $T_{\rm MB}^{\rm HI}$ (in the unit of Kelvin) denotes the brightness temperature of \ion{H}{1} emission, and  $dv$ is the spectral resolution in the unit of km s$^{-1}$. Contribution from $\rm H_2$ emission is neglected as the foreground is overwhelmingly dominated by \ion{H}{1} emission. The hydrogen column density of W44 can be obtained in a similar way using the the CO-$\rm H_2$ conversion factor $X_{\rm CO} = 2\times10^{20}$ \text{$\rm{cm^{-2}}$ / (K km s$^{-1}$)} in giant molecular clouds (\citealt{2011MNRAS.418..664N}) :
\begin{equation}
    \begin{aligned}
        N{\rm _H^{W}}&=N_{\rm H I}+2N_{\rm H_2}\approx 2N_{\rm H_2}= 2X_{\rm CO}\int_{v_{\rm min}}^{v_{\rm max}} T_{\rm MB}^{\rm CO} dv
        \end{aligned}
        \end{equation}
where $T_{\rm MB}^{\rm CO}$ (in the unit of Kelvin) is the brightness temperature of CO emission. On contrast to Eq.~\ref{eq.10}, contribution of \ion{H}{1} in molecular cloud is neglected as $\rm H_2$ emission dominates here. By adopting a linear relation between the hydrogen column density and optical extinction $A_{\rm v}$ \citep{2009MNRAS.400.2050G}:
\begin{equation}
    N_{\rm H}=(2.21\pm0.09)\times10^{21}A_{\rm v}
\end{equation}
we obtained the $A_{\rm v}$({\rm mag}) contributed by the W44 and the foreground/background, respectively. Fig.~\ref{fig.colden_HI} and Fig.~\ref{fig.colden_CO} display the results of such calculations. 

Fig.~\ref{fig.colden_HI} quantitatively verifies the discussion about the foreground above. $A_{\rm v}$ shows the peak significance in the top-north corner, where the \ion{H}{I} emission reaches the peak intensity, while in the W44 SNR regions and south-bottom region $A_{\rm v}$ reaches the minimum,  cohering to the pattern discovered earlier in Fig.~\ref{fig.HI-Planck}. The $A_{\rm v}$-AM histogram also exhibits the positive correlation between the two variables that a large $A_{\rm v}$ value corresponds to a large AM value as well, i.e. in regions where the \ion{H}{I} emission is intense, the agreement between Planck polarization and VGT measurement tend to be better. It implies a significant contribution from the foreground to the disagreement seen in Fig.~\ref{fig.vgt}.

Fig.~\ref{fig.colden_CO} shows the extinction map towards the SNR W44 obtained from the CO emission line. Around the filamentary structure, $A_{\rm v}$ displays the greatest magnitude $\ge1.00$ mag, which is greater than the foreground's $A_{\rm v}\le0.75$ mag. It indicates the dominance of W44's contribution in Planck so that we have a high consistency of the VGT-Planck. In the misalignment regions ($\rm AM\le0$, west-north corners), $A_{\rm v}$ drops to $\sim0.75$ mag and the foreground's $A_{\rm v}$ increases to $\sim0.8$ mag. Therefore, the foreground has more contribution to the Planck in the region. The histogram similarly reveals the proportional correlation between $A_{\rm v}$ and AM. The graphs thus confirm the association between the foreground interference and the VGT-Planck inconsistency.

\subsection{Magnetic field for various velocity components}
We adopt SCOUSEPY to decompose the W44 emission line into various components. The decomposed spectrum is further processed by a hierarchical agglomerative clustering technique ACORNS (Agglomerative Clustering for ORganising Nested Structures) \footnote{\url{https://github.com/jdhen
shaw/acorns.}}. We perform the ACORNS clustering only on the most robust spectral velocity components decomposed by SCOUSEPY, i.e. (i) the signal-to-noise ratio is greater than three; (ii) the minimum radius of a cluster to be 30$''$, which is 150\% of the beam resolution; (iii) for two data points to be classified as "linked", the absolute difference in measured velocity dispersion can be no greater 1.0 km/s.  The clustering comprises $\sim97\%$ of all data and results in a number of hierarchical and non-hierarchical components. Considering the fact that both components contribute to VGT measurement, we further combine them into the four most significant velocity components based on the un-decomposed averaged spectral lines (see Fig.~\ref{fig.intensity}).

As shown in Fig.~\ref{fig.decomposed-spectrum}, our final decomposed spectrum clusters are dominated by 4 components. Panel b in Fig.~\ref{fig.decomposed-spectrum} shows the spectrum of the four major components. The blue curve with the highest intensity peak shows the dominating component 1, with three other components that are less significant. Note that because the velocity difference between each component is greater than their velocity dispersion, we expect these four components to be physical structures in real space. 


Fig.~\ref{fig.velocity-components} displays the individual images of the velocity-wise magnetic field components. In general all components exhibit a magnetic field orientation along the northeast direction. Component 1 (at $v\sim45$ km s$^{-1}$) exhibits the largest coverage of the filamentary SNR structures shown in Fig.~\ref{fig.vgt}, corresponding to the most prominent intensity distribution in the Fig.~\ref{fig.decomposed-spectrum} spectrum. The magnetic field orientations visually cohere with the overall magnetic field map in Fig.~\ref{fig.vgt} in a highly consistent manner as well, indicating its dominating effect on the whole picture. It also shows agreement with Planck polarization at high-intensity regions. This can be easily understood based on the fact that component 1 has the most significant intensity so that its contribution dominates the projected magnetic field map, which is contributed by all components.

The other three components have a smaller contribution to the SNR magnetic field overall, yet they reveal several details that are overwhelmed by the major regions with high intensities in component 1. Component 2 (at $v\sim80$ km s$^{-1}$) and 4 (at $v\sim50$ km s$^{-1}$) both indicate novel SNR structures that were not discovered initially in the complete image of the W44 magnetic field in Fig.~\ref{fig.vgt}. At the center of component 2 exists a small-scale SNR structure, which is an ellipse-like density structure and a roundabout magnetic field are situated at. In the Northeast corner of component 4, a SNR structure stretched along the magnetic field in the vertical direction is discovered. These SNR structures were unseen in the same regions in component 1, as their low intensities were overshadowed by the major high-intensity structures in the overall picture. A precise investigation for such concealed parts may provide useful information towards the study of the structural evolution of SNRs.


\section{Discussion} 
\subsection{The range of VGT applications}

The VGT has been successfully used to study magnetic fields in diffuse H I \citep{2017ApJ...837L..24Y,2018MNRAS.480.1333H,2019ApJ...886...17H,2020ApJ...888...96H,2020MNRAS.496.2868L}, molecular clouds, both in the galactic disk \citep{2019NatAs...3..776H,2019ApJ...884..137H,2021ApJ...912....2H,2020arXiv200715344A} and the Central Molecular Zone (CMZ; \citealt{2021arXiv210503605H}). This work shows that the VGT provides a useful tool for studying magnetic fields in supernovae remnants. The latter exhibit both shock regular compression and turbulence and therefore it is obvious from the very beginning whether the VGT could be accurate for them. However, our study has proven a good tracing power of the VGT in this setting. 

The present work also boosts our confidence in magnetic fields using other branches of the Gradient Technique, for instance, the branches utilizing the synchrotron intensity gradient (SIGs; \citealt{2018ApJ...855...72L}) and synchrotron polarization gradients (SPGs; \citealt{2018ApJ...865...59L}), as well as intensity gradients (IGs; \citealt{2019ApJ...886...17H}). The theoretical justification of these techniques is similar to that of the VGT and it is rooted in the theory of MHD turbulence and turbulent reconnection. In some cases, e.g. while studying the magnetic fields in clusters of galaxies as in \cite{2020ApJ...901..162H} we do not have the corresponding polarization data to compare with. The present study shows that the measured gradients arise mostly from turbulence and therefore can represent magnetic fields correctly in a wide variety of astrophysical conditions.  

\subsection{Prospects of the VGT}
Probing magnetic fields in supernovae remnants (SNRs) is an essential yet technically challenging topic. Our research towards the SNR W44 magnetic field with the Velocity Gradients Technique (VGT; \citealt{2017ApJ...835...41G,2017ApJ...837L..24Y,2018ApJ...853...96L,2018MNRAS.480.1333H}) has led to several intriguing discoveries. Quantitative studies verify the credibility of the VGT on supernovae investigations. By comparing the W44's magnetic field probed with the VGT and the Planck 353 GHz dust polarization, shown in Fig.~\ref{fig.vgt}, we preliminary confirm the high consistency on the measurements by the novel VGT and the commonly-recognized Planck polarization. In other regions with a less intense gas emission, we obtain evidence, in Fig.~\ref{fig.HI-Planck}, Fig.~\ref{fig.colden_HI}, Fig.~\ref{fig.colden_CO}, indicating that the VGT-Planck consistency is affected by the foreground interference on the Planck polarization data. The advantages of the VGT in removing the foreground interference have once again been highlighted as in previous research (\citealt{2020ApJ...888...96H,2020MNRAS.496.2868L}). Furthermore, grouping the VGT-probed results intensity-wise as such helps to specify the most ideal conditions for applying the novel technique in SNR magnetic field tracing.

Meanwhile, compared to the general magnetic field probing approach like the Planck dust polarization, the VGT possesses multiple advantages in tracing magnetic fields in SNRs. Along with the capability of removing the foreground interference as stated above, the technique also exhibits potentials in decomposing the SNR magnetic field into various components with different velocities and performing in-depth studies towards the components individually. The W44 velocity-wise magnetic field components unfold noteworthy details. Individual investigations on such components provide valuable insights about certain SNR structures that were completely buried under the prominent high-intensity regions in the initial magnetic field tomography. We reasonably expect that more useful information about the SNR will be found by such a novel perspective of research. The significance of this potential should not be underestimated, as so far the VGT remains the sole magnetic field tracing technique that is capable of identifying such informative details. 

\subsection{Prospects for cosmic rays' study}
The diffusion of cosmic rays (CRs) is a fundamental ingredient of astrophysics and has a broad impact on diverse astrophysical problems. Theoretically, the recent development of MHD turbulence theories since GS95 and LV99 has brought a significant shift to the physical picture of CR diffusion (\citealt{2002PhRvL..89B1102Y,2004ApJ...614..757Y, 2008ApJ...673..942Y, 2013ApJ...779..140X,2014ApJ...784...38L,2021arXiv211115066H}). It reveals that the CR diffusion and the diffusion coefficient have a strong dependence on the properties of MHD turbulence, i.e., the sonic Mach number $M_{\rm S}$ and Alfv\'en Mach number $M_{\rm A}$. Consequently, in the molecular cloud that SNRs interact with, the properties of the highly compressive and also magnetized turbulence determine the diffusion of CRs that are accelerated by SNRs.

The two important quantities $M_{\rm A}$ and $M_{\rm S}$ are also accessible by VGT. The dispersion of velocity gradient's orientation reveals $M_{\rm A}$ (\citealt{2018ApJ...865...46L}) and the amplitude of velocity gradient outputs $M_{\rm S}$ (\citealt{2020ApJ...898...65Y}). The application of VGT provides an important observational test of fundamental theories of CR diffusion.

\section{Conclusion} 

Tracing 3D magnetic fields in the molecular cloud that SNRs interact with is still obscured. In this work, we target the SNR W44 to test the VGT's applicability in probing 3D magnetic fields. We employ several data sets, including CO, $\rm HCO^+$, and H I emission lines, as well as dust polarization data from Planck 353 GHz measurement. Our main discoveries are:
\begin{enumerate}
    \item The SNR magnetic fields probed with the VGT generally agree well with the results acquired with the Planck dust polarization, especially in the high molecular gas emission intensity regions.
    \item Using GALFA-HI data we evaluate the effect of the foreground on the Planck polarization. We find that the discrepancy between the Planck and the VGT arises due to the contribution of the foreground. Therefore, unlike Planck polarization, the VGT provides a better measure of the magnetic field in the supernovae remnant. 
    \item We decompose the SNR W44 into four individual components at various velocities using the Python implement SCOUSEPY and use VGT to trace the magnetic field associated with different components.
    \item We show that W44's velocity component 1 at $v\sim45$ km s$^{-1}$ exhibits the largest coverage of the filamentary SNR structures and gives the best agreement with Planck polarization in terms of magnetic field orientation.
    \item We verify the VGT's potential in tracing the magnetic fields of the decomposed individual velocity components. We show the ability of VGT to measure three-dimensional magnetic field tomography. 
\end{enumerate}
The study verifies the credibility of the VGT on the SNR magnetic field tracing and reveals the potential of the novel technique for highly disturbed regions. It also shows prospects for the technique for future application including tomographic studies of the SNRs. We expect more profound findings to be revealed using the novel Velocity Gradients Technique.

\section*{Acknowledgements}
M.L. acknowledges the invaluable support and accompany by Liu's family and close acquaintances. Y.H. acknowledges the support of the NASA TCAN 144AAG1967. A.L. acknowledges the support of the NSF grant AST 1715754 and NASA ATP AAH7546. We acknowledge the Nobeyama Observatory for providing the data of the W44 region.

\section*{Data Availability}
The data underlying this article will be shared on reasonable
request to the corresponding author.



\bibliographystyle{mnras}
\bibliography{reference} 








\bsp	
\label{lastpage}
\end{document}